%% file: main.tex
\documentclass[sigconf, screen]{acmart}

\usepackage{xcolor}
\usepackage{array}
\usepackage{pifont}
\definecolor{commentgreen}{RGB}{9,136,66}

\startPage{1}
\usepackage{algorithm}
\usepackage{algpseudocode}
\usepackage{graphicx} 
\usepackage{booktabs} 
\usepackage{array}
\usepackage{tabularx}
\usepackage{subcaption}
\usepackage{enumitem}

\copyrightyear{2025}
\acmYear{2025}
\setcopyright{acmlicensed}\acmConference[ICS '25]{2025 International Conference on Supercomputing}{June 8--11, 2025}{Salt Lake City, UT, USA}
\acmBooktitle{2025 International Conference on Supercomputing (ICS '25), June 8--11, 2025, Salt Lake City, UT, USA}
\acmDOI{10.1145/3721145.3725746}
\acmISBN{979-8-4007-1537-2/2025/06}

\begin{document}

\title{CB-SpMV:A Data Aggregating and Balance Algorithm for \underline{C}ache-Friendly \underline{B}lock-Based \underline{SpMV} on GPUs}

\author{Xing Cong}
\email{congxing@buaa.edu.cn}
\affiliation{
  {School of Computer Science and Engineering}\\
  {Beihang University}
  {Beijing}\\
   \country{China}
}

\author{Fukai Sun}
\email{sunfukai@buaa.edu.cn}
\affiliation{
 {School of Computer Science and Engineering}\\
  {Beihang University}
  {Beijing}\\
  \country{China}
}

\author{Yifan Chen}
\email{chenyifan@buaa.edu.cn}
\affiliation{
  {School of Computer Science and Engineering}\\
  {Beihang University}
  {Beijing}\\
   \country{China}
}

\author{Chenhao Xie}
\email{xiechenhao@buaa.edu.cn}
\authornote{Corresponding author}
\affiliation{
 {School of Computer Science and Engineering}\\
  {Beihang University}
  {Beijing}\\
   \country{China}
}

\author{Yi Liu}
\email{yi.liu@buaa.edu.cn}
\affiliation{
  {School of Computer Science and Engineering}\\
  {Beihang University}
  {Beijing}\\
   \country{China}
}

\author{Depei Qian}
\email{depeiq@buaa.edu.cn}
\affiliation{
  {School of Computer Science and Engineering}\\
  {Beihang University}
  {Beijing}\\
   \country{China}
}

\begin{abstract}
Sparse matrix-vector multiplication (SpMV) is crucial in computational science, engineering, and machine learning. Despite substantial efforts to improve SpMV performance on GPUs through various techniques, issues related to data locality, hardware utilization, and load balancing persist, leaving room for further optimization.
This paper presents CB-SpMV, a cache-friendly SpMV optimization algorithm, using a novel data convergent and adaptable 2D blocking structure. The matrix in CB-SpMV is divided into independent sub-blocks, with virtual pointers aggregating different types of intra-block data for better cache-level data locality. To enhance hardware utilization, a block-aware column aggregation strategy and the selection of sub-block formats are proposed to accelerate computation and adapt to varying sparse matrices. Finally, an inter-block load-balancing algorithm is designed to ensure efficient workload distribution across thread blocks.
Experimental evaluations on 2,843 matrices from the SuiteSparse Collection show that CB-SpMV significantly improves cache hit rates and achieves average speedups of up to 3.95× over state-of-the-art methods like cuSPARSE-BSR, TileSpMV, and DASP on NVIDIA A100 and RTX 4090 GPUs. The implementation is available at: \url{https://github.com/xing-cong/CB-Sparse}.
\end{abstract}



\maketitle 

\protect\input{./sections/introduction}

\protect\input{./sections/motivation}

\protect\input{./sections/method}

\protect\input{./sections/experiment}

\protect\input{./sections/related_work}

\protect\input{./sections/conclusion}


\begin{acks}
The work is supported by the Natural Key Research and Development Program of China (2023YFB3002902) and National Natural Science Foundation of China (No.62322201 and U23B2020). 
\end{acks}


\bibliographystyle{ACM-Reference-Format}
\bibliography{sample-base}


\end{document}

%% file: sections/introduction.tex
\section{Introduction}

\label{section:1}

Sparse matrix operations, a cornerstone of computational science and engineering, are essential in numerical simulations, data analysis, and machine learning. Sparse matrices, characterized by irregularly distributed non-zero elements, pose challenges such as poor memory locality and load imbalances in parallel computing, which hinder computational efficiency. Among these operations, sparse matrix-vector multiplication (SpMV) is one of the most fundamental and widely studied kernels. Research has focused on optimizing memory access through novel storage formats (e.g., CSR5\cite{liu2015csr5}, LSRB-CSR\cite{liu2015lsrb}, TileSpMV\cite{niu2021tilespmv}), dynamic storage format selection(e.g., Alphasparse \cite{du2022alphasparse}), and leveraging Tensor Core hardware(e.g., DASP\cite{lu2023dasp}). Despite these advancements, issues related to cache efficiency, hardware utilization, and load balancing persist, leaving room for further optimization. 

\begin{figure*}[t!]
    \centering
    \includegraphics[width=1\linewidth]{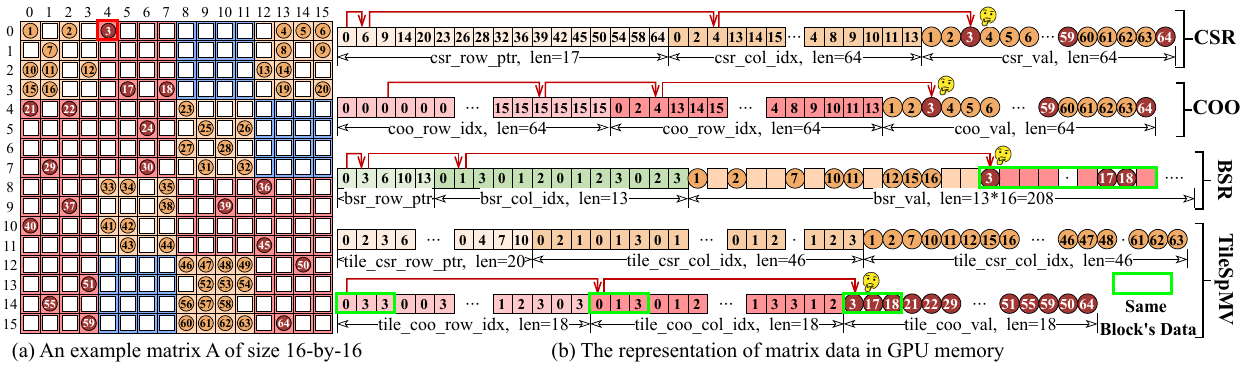}
    \vspace{-20pt}
    \caption{\small{\textbf{(a) illustrates a 16×16 example matrix divided into 4×4 sub-blocks, with 13 non-zero sub-blocks highlighted. (b) depicts the data layout of the matrix in GPU global memory across four sparse storage formats (CSR, COO, BSR, TileSpMV), along with the memory access patterns for retrieving the third non-zero element located at (0,4) in the example matrix.}}}
    \vspace{-10pt}
    \label{fig:1}
\end{figure*}

\textbf{Regarding data locality}, both the widely studied CSR format and the state-of-the-art block-based format\cite{ji2022tilespmspv,niu2022tilespgemm,niu2021tilespmv} store the coordinate array and value array separately. This design necessitates frequent access to the coordinate array during SpMV operations, which involves large memory jumps to retrieve elements from the value array. Such a skipping storage structure significantly reduces the hit rates of the GPU’s L1 and L2 caches. A detailed analysis of this issue is provided in the following section \ref{section:2.2}.

\textbf{Regarding hardware utilization}, although previous research works\cite{niu2021tilespmv} introduce block-based SpMV for data locality and a variety of formats to handle sub-blocks with different sparsity levels, it still suffers from inadequate hardware utilization on GPUs in the case of highly sparse sub-blocks. For instance, when a sub-block contains few non-zero elements, the GPU still needs a warp, the basic thread unit of GPUs, to process, which leads to most of the warp threads remaining idle. Such a similar situation also exists for Tensor Core accelerated SpMV\cite{lu2023dasp}. In the data layout of DASP, the 16×8 TCU(Tensor Core Unit) produces 256 values per cycle, but only the 16 diagonal values are relevant for SpMV, limiting the tensor core’s effective utilization.

\textbf{Regarding load balance}, block-based methods usually process each sub-block at the warp level. However, the number of non-zero elements within each sub-block varies significantly. Consequently, the total number of non-zero elements that each thread block (generally composed of eight warps \cite{cook2012cuda}) needs to process can differ substantially. This imbalance results in uneven workloads across streaming multiprocessors (SMs), ultimately leading to performance degradation.
To accelerate SpMV calculation by improving the data locality, hardware utilization, and load balance simultaneously, this paper proposes Cache-friendly Block-based SpMV, \textbf{CB-SpMV}, including a novel data convergent and adaptable 2D block structure and a series of optimization policies: \textbf{Intra-Block Data Aggregating} to unite the different storage data formats within the sub-block; \textbf{Computation Adaptation} for handling blocks with varying sparsity, and \textbf{TB(Thread Bolck)-Load-Balance} to ensure equitable workload distribution across thread blocks.

The design of CB-SpMV stems from a key insight: after partitioning a matrix into sub-blocks, the data within each sub-block is independent and self-contained, with the coordinates of non-zero elements relative to the sub-block itself. Leveraging this property, CB-SpMV was developed to treat each sub-block as an independent unit, compactly storing the different types of coordinates and non-zero values via format uniting and an efficient virtual pointer structure. This design significantly reduces scattered memory access, enhancing data locality and improving L1 and L2 cache hit rates. However, achieving optimal performance with CB-SpMV introduces additional challenges, particularly in maintaining block independence while addressing issues such as load balancing and the trade-off between parallelism and utilization.

To address the hardware utilization challenge for varying sparsity levels, we apply a block-aware column aggregation strategy for sparse sub-blocks and choose either CSR or Dense formats for denser sub-blocks, optimizing computation efficiency. Additionally, a parallel load balancing algorithm leveraging priority queues ensures equitable workload distribution, further enhancing performance. 
Experimental evaluations on 2,843 matrices from SuiteSparse demonstrate the effectiveness of CB-SpMV. On the RTX 4090 GPU, the method increases L1 and L2 cache hit rates by 82\% and 19\%, respectively, compared to TileSpMV\cite{niu2021tilespmv}, and achieves improvements of 15.62× and 10.05× over cuSPARSE-BSR\cite{cusparse}. It also delivers speedups of 2.95×, 3.06×, and 2.76× over cuSPARSE-BSR, TileSpMV, and DASP\cite{lu2023dasp}. The contributions of this paper are summarized as follows:

\vspace{-3pt}
\begin{itemize}[left=0pt]
    \item \textbf{2D Blocking Structure:} We introduce an innovative 2D blocking structure whose design enhances sub-block independence and enables faster, more convenient sub-block mapping to warps.
    \item \textbf{Intra-Block Data Aggregation:} We propose a data aggregating method using a virtual pointer structure to unite the format of different data within the sub-block.
    \item \textbf{Computation Adaptation Strategy:} To address varying sparsity levels among sub-blocks, we propose a computation adaptation strategy that optimizes hardware utilization and improves parallel efficiency.
    \item \textbf{Thread-block Load Balancing Optimization:} We develop a parallel load balancing algorithm to mitigate workload disparities among thread blocks. 
    \item \textbf{CB-SpMV Framework:} Combining them all, we propose CB-SpMV, a cache-friendly block-based framework for SpMV on GPUs, leveraging data aggregation and balancing to boost performance. Experiments on two GPUs demonstrate that CB-SpMV achieves higher cache hit rates and outperforms SOTA methods.
\end{itemize}
\vspace{-3pt}

The remainder of this paper is organized as follows: Section~\ref{section:2} describes the background of the sparse format and SpMV, with a comprehensive discussion of its data locality, hardware utilization, and load balance challenges. The proposed CB-SpMV and relative optimization method are presented in Section~\ref{section:3}. Section~\ref{section:4} presents the evaluation results of the CB-SpMV on RTX 4090 and A100. We discuss the related work in Section~\ref{section:5} and conclude this paper in Section~\ref{section:6}.

%% file: sections/motivation.tex
\section{Background, Motivation and Challenge}
\label{section:2}

\subsection{Sparse Matrix Storage Format}

Sparse matrices, characterized by a few non-zero elements per row, require specialized storage formats to improve access efficiency. The COO format is widely used for its simplicity and compatibility with data storage, stores non-zero elements as \texttt{(row\_idx, col\_idx, coo\_val)} triplets, enabling rapid construction and easy conversion to other formats. The CSR format, among the most widely adopted, organizes data into \texttt{row\_ptr}, \texttt{col\_idx}, and \texttt{csr\_val} arrays, offering compact storage and efficient row-wise operations. With the rise of block accelerators such as Tensor Cores, block-based formats like the Block Compressed Sparse Row (BSR) format have gained traction. BSR partitions matrices into fixed-size blocks, using block-level arrays (\texttt{blk\_row\_ptr}, \texttt{blk\_col\_idx}) and storing block data in \texttt{bsr\_val}, including both non-zero and zero elements. This structure is particularly effective for dense matrices. Advanced methods such as TileSpMV\cite{niu2021tilespmv} further optimize block-based SpMV by employing mixed storage formats at the block level to reduce zero storage overhead, as illustrated in Fig.\ref{fig:1}.

\subsection{SpMV and Motivation}
\label{section:2.2}

\begin{figure}[t]
    \centering
    \includegraphics[width=1\linewidth]{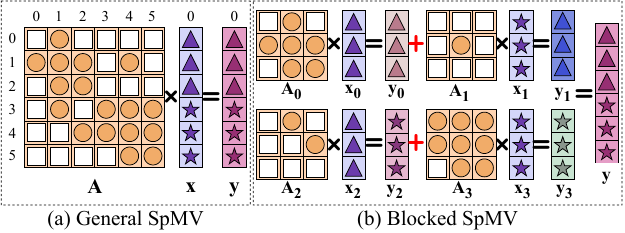}
    \caption{\small{\textbf{An example of an SpMV that multiplies a 6-by-6 sparse matrix $A$ by a vector $x$ to get a vector $y$.}}}
    \label{fig:2}
\end{figure}

SpMV calculates the product of a sparse matrix $A$ and a dense vector $x$, iterating over non-zero elements to update the output vector $y$. As a core operation in sparse linear algebra, SpMV is widely used in scientific computing and engineering applications as a core operation in sparse linear algebra. Fig.\ref{fig:2} and Alg.\ref{alg:1} illustrate its computation and logic.

\begin{algorithm}
\caption{A pseudocode of parallel CSR SpMV.}
\small
\begin{algorithmic}[1]
	\State $s\_x \gets x$ \Comment{$s\_x$ is in shared memory}
	\For{$i=0$ to $m$ \textbf{in parallel} }
	    \State $sum \gets 0$
	    \For{ $j=row\_ptr[i]$ to $row\_ptr[i+1]$ }
                \State \textcolor{commentgreen}{//$m+1$ elements are spanned from $row\_ptr$ to $col\_idx$}
		    \State $val\_x \gets s\_x[col\_idx[j]]$ 
                \State \textcolor{commentgreen}{//$nnz$ elements are spanned from $col\_idx$ to $csr\_val$}
		    \State $sum \gets sum + val\_x \times csr\_val[j]$
		\EndFor
		\State $y[i] \gets sum$
	\EndFor
\end{algorithmic}
\label{alg:1}
\end{algorithm}

However, performing SpMV with the CSR format introduces significant data locality issues. As shown in Fig.\ref{fig:1}, accessing the element at (0,4) and other elements begins with querying \texttt{row\_ptr} to determine the starting positions of non-zero elements, which exhibits high locality due to adjacent data. However, subsequent accesses to \texttt{col\_idx} and \texttt{csr\_val} involve significant memory jumps, leading to a marked reduction in cache efficiency. These jumps recur within rows, further degrading data locality and overall cache performance. Unlike CPUs, which rely on hardware prefetchers to mitigate memory latency\cite{guo2024camlb}, GPUs depend on high parallelism to hide latency, offering limited prefetching capabilities. For example, on an RTX 4090 GPU with 128 SMs, each with 128 KB of L1 cache, the theoretical L1 cache per thread is only 64B when scheduling multiple TBs per SM. Similarly, the shared 72MB L2 cache is insufficient to accommodate large memory spans, making it challenging to leverage GPU caches effectively. Other formats like COO and BSR also suffer from locality challenges. COO incurs higher jumps due to the direct traversal of non-zero elements. At the same time, BSR improves locality by processing data in blocks but at the cost of storing zero elements, reducing efficiency for sparse matrices. TileSpMV addresses some of BSR’s limitations by compressing sub-blocks but fails to fully exploit the inherent locality of BSR-dense sub-blocks, reintroducing issues similar to CSR, as shown in Fig.\ref{fig:1}. These limitations motivate the design of a novel sparse storage structure that preserves BSR’s locality advantages while eliminating zero-element storage, enabling more efficient SpMV computation.

\begin{figure}[t!]
    \centering
    \includegraphics[width=1\linewidth]{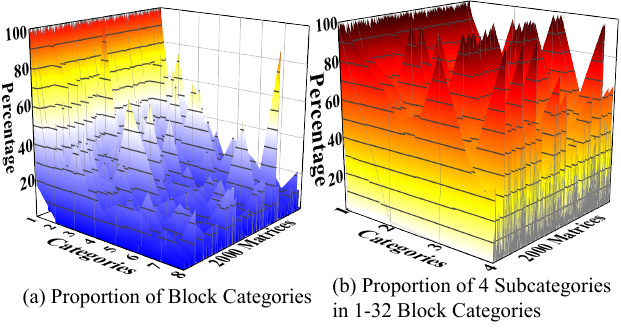}
    \vspace{-20pt}
    \caption{\small{\textbf{Distribution of non-zero elements in 2000 matrices (with over 10,000 non-zero elements) under 16×16 block partition: (a) Proportion of sub-blocks across eight categories (1-32, 33-64, …, 225-256); (b) Further subdivision of the 1-32 category into four subcategories (1-8, 9-16, 17-24, 25-32).}}}
    \label{fig:3}
    \vspace{-10pt}
\end{figure}

Besides the data locality issue,  we also observe that hardware under-utilization and load imbalance limit the efficiency of block-based SpMV. Regarding the hardware under-utilization, the BSR format faces a critical limitation due to the significant variation in sub-block sparsity after partitioning. For example, Nvidia GPUs comprise 32 threads in a wrap. If the sub-blocks contain only a few non-zero elements, most threads will be idle, wasting computational resources. Fig.\ref{fig:3}(a) illustrates the distribution of non-zero elements in sub-blocks for 2000 SuiteSparse matrices (each with over 10,000 non-zero elements) using a 16×16 block size. Sub-blocks were categorized into eight ranges (1–32, 33–64, …, 225–256), with over 90\% of sub-blocks in most matrices falling within the 1–32 range, averaging 81.89\% across all matrices. In Fig.\ref{fig:3}(b), the 1–32 range is further divided into four subcategories, revealing that sub-blocks with 1–8 non-zero elements constitute 59.36\% of the total, and those with 9–16 elements account for 20.35\%. This indicates that 59.36\% of sub-blocks have a warp-level thread utilization below 75\%, while 20.35\% falls below 50\%, leading to significant efficiency losses. Although TileSpMV\cite{niu2021tilespmv} partially addresses this by consolidating sparse sub-blocks and processing them with CSR5\cite{liu2015csr5} while handling the rest with TileSpMV, this approach incurs additional overhead from merging and consolidation. Hence, more efficient solutions to this issue are still needed.

\begin{figure}[t!]
    \centering
    \includegraphics[width=1\linewidth]{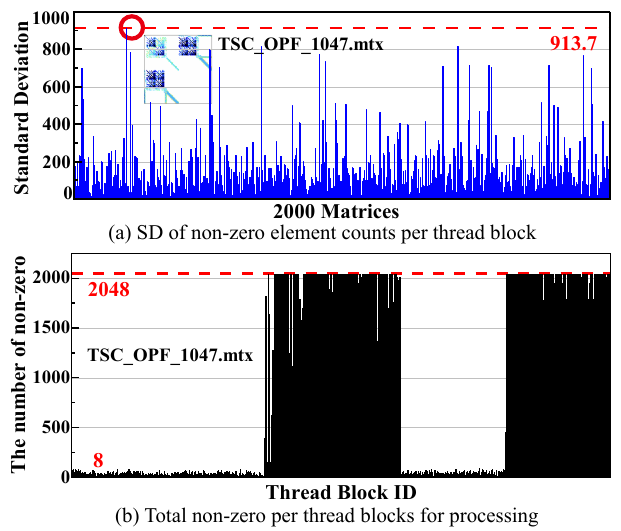}
    \caption{\small{\textbf{Standard deviation and distribution of non-zero elements per thread block: (a) Standard deviation across 2000 matrices; (b) Distribution for a specific matrix (\texttt{TSC\_OPF\_1047.mtx}).}}}
    \vspace{-10pt}
    \label{fig:4}
\end{figure}

Regarding load imbalance, thread blocks(TBs) serve as the basic scheduling units for SMs in GPU computation, with each block comprising multiple warps. Due to varying sparsity among sub-blocks, the number of non-zero elements assigned to each thread block often differs significantly. Fig.\ref{fig:4} highlights this imbalance across 2000 matrices (over 10,000 non-zero elements each) with a 16×16 block size. The standard deviation of non-zero elements per thread block varies widely, peaking at 913.7 for \texttt{TSC\_OPF\_1047.mtx}, indicating severe load imbalance. Since GPU scheduling assigns thread blocks to SMs in a round-robin manner, the execution time for an SM is dictated by its slowest thread block, typically the one with the heaviest load. This imbalance in workload distribution among thread blocks can significantly degrade performance without proper mitigation.

\subsection{Software and Hardware Constraints}

We further observe that the programming model and GPU itself raise significant constraints when designing mixed precision and better data locality block-based SpMV. The constraints can be categorized into three aspects. First, GPU functions are designed to process homogeneous data that contracts with sparse structures. Taking \texttt{memcpy()} and \texttt{malloc()} as an example, only homogeneous data types are supported during data transferring. However, sub-block data often include mixed types and precisions—e.g., \texttt{int} for coordinates and \texttt{float} or \texttt{double} for numerical values—making it impossible to transfer these data types simultaneously, which incurs high data management overhead. Second, the GPU’s memory alignment mechanism also introduces organizational challenges. Different data types, such as \texttt{int}(4B) and \texttt{double}(8B), may not align contiguously in memory, resulting in inefficient storage and access patterns. Due to these two constraints, although the state-of-the-art block-based structures partition large matrices into sub-blocks, they cannot fully capture the data locality within sub-blocks. Moreover, the parallel executed model poses significant difficulties in mapping imbalanced computing tasks to GPU hardware. For example, Nvidia GPUs process threads within warps in a SIMD manner and group multiple warps to form a thread block for resource scheduling. Since each thread runs the same code and is scheduled simultaneously by a single thread block, it is difficult to achieve highly flexible task assignments for high hardware utilization and load balance.

Thus, by considering only the block-based SpMV itself without breaking these software and hardware constraints, new bottlenecks will emerge in this kind of SpMV.

%% file: sections/method.tex
\section{CB-SpMV}

\label{section:3}

To overcome the challenges of block-based SpMV, we propose a cache-friendly block-based SpMV approach, CB-SpMV, specifically designed for GPUs. This approach incorporates a novel mixed-precision method and an adaptable 2D sparse structure tailored to matrices with varying sparsity. In our design, CB-SpMV efficiently transforms the input sparse matrix into the proposed high-locality 2D sparse structure during the data loading phase using innovative data aggregation and balancing algorithms. To fully leverage the benefits of the CB-SpMV format, we also redesign the computation logic of the SpMV kernel. Notably, the threads within each block are assigned in the original manner to ensure the high level of parallelism inherent to GPUs is preserved.

\begin{figure}[t!]
    \centering
    \includegraphics[width=1\linewidth]{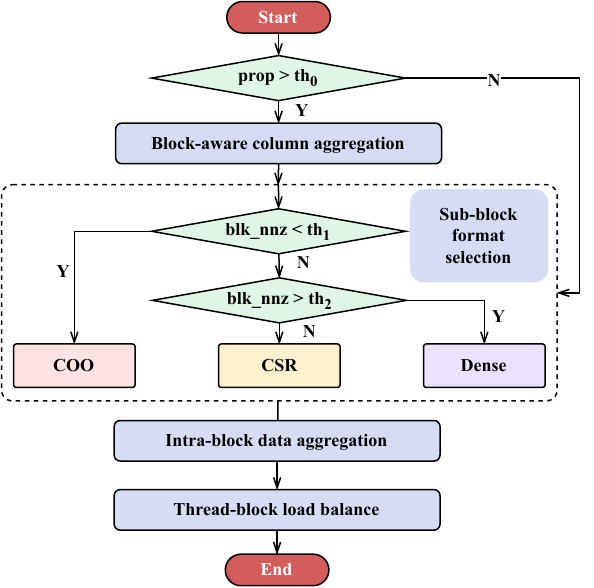}
    \caption{\small{\textbf{The overview flow chart of CB-SpMV}}}
    \label{fig:7}
\end{figure}

\begin{figure*}[t]
    \centering
    \includegraphics[width=1\linewidth]{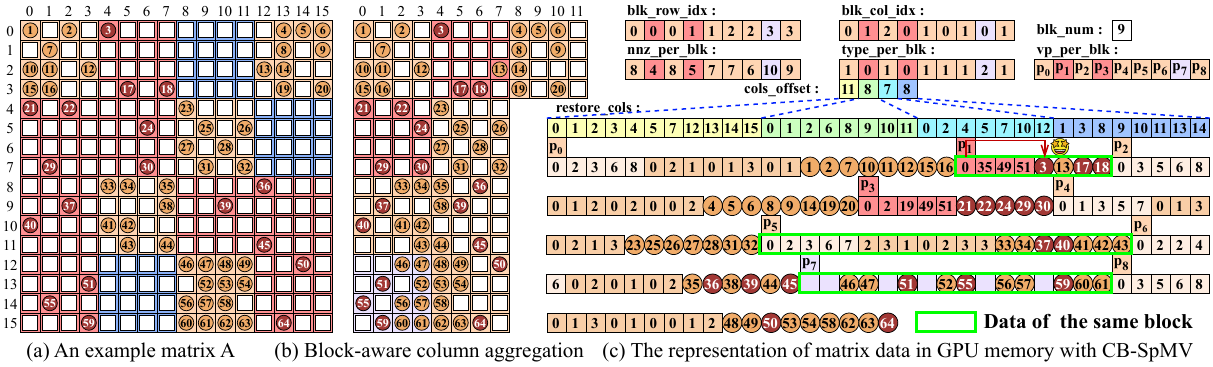}
    \caption{\small{\textbf{The storage format and memory organization of CB-SpMV. (a) illustrates a 16-by-16 example matrix, where different colors indicate that sub-blocks adopt distinct storage formats. (b) depicts the matrix after column aggregation. (c) shows the 2D metadata of CB-SpMV. In comparison, CB-SpMV aggregates different types of data within the sub-block for better data locality.}}}
    \label{fig:5}
\end{figure*}

As shown in Fig.\ref{fig:7}, the overview flow chart of CB-SpMV incorporates three main components: an intra-block data aggregation to enhance data locality, a block-aware column aggregation, and an efficient inter-block load balancing algorithm to improve warp and block-level parallelism as well as hardware utilization. First, input data is loaded as a block-based COO format, similar to HiCOO \cite{li2018hicoo}. After checking the characteristics of the input matrix, the input data passes through the column aggregation to increase non-zero value density. Then, the compressed matrix is transformed into a 2D sparse structure, and the sub-block format is selected for efficient processing. After that, the various types of data within sub-blocks are aggregated to capture the data locality. Finally, the load balancing is achieved via an inter-block exchange algorithm, ensuring an even distribution of non-zero elements among thread blocks, ultimately improving computational efficiency. We set different thresholds to convert the sparse matrix for optimal data locality and parallelism, with the detailed threshold setting method introduced alongside each component.\enlargethispage{-12pt}

\subsection{2D Sparse Structure}

In this work, CB-SpMV divides the input sparse matrix into uniform 16×16 sub-blocks to balance data locality, parallelism, and thread utilization, which serve as the fundamental computational units, with each sub-block mapped to a warp for intra-block computation. This block size ensures efficient warp-level execution on NVIDIA GPUs, avoiding under-utilization from sparse sub-blocks or reduced parallelism from overly large blocks. The method employs a two-level metadata structure: a high-level structure for locating sub-blocks and a low-level structure for managing data within each sub-block. Fig.\ref{fig:5} illustrates this 2D sparse structure. 

The high-level metadata adopts the COO format for efficient localization of sub-block positions and warp binding. It consists of five arrays: \texttt{blk\_row\_idx} and \texttt{blk\_col\_idx} for row and column indices of non-zero sub-blocks, \texttt{nnz\_per\_blk} for the number of non-zero elements in each sub-block, \texttt{vp\_per\_blk} for the starting GPU memory addresses of sub-block data, and \texttt{type\_per\_blk} to specify the storage format of each sub-block. The low-level structure defaults to the COO format for rapid conversion from the input data to the CB-SpMV structure. To treat the different types of data within the sub-blocks as a united structure, these sub-blocks are then aggregated into a single structure array (\texttt{blk\_data}) on the CPU and packed contiguously into a one-dimensional meta-data (\texttt{mtx\_data}) on the GPU via intra-block data aggregation.

\begin{figure}[t!]
    \centering
    \includegraphics[width=1\linewidth]{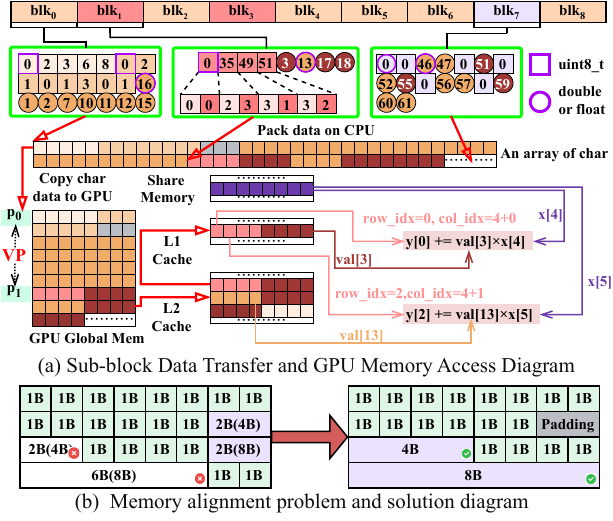}
    \caption{\small{\textbf{Sub-block Data Transfer and GPU Memory Access: (a) Packaging of sub-block data on the CPU, transfer to GPU global memory, and access via L1, L2 caches, and shared memory; (b) Memory alignment issue and its resolution.}}}
    \label{fig:6}
    \vspace{-15pt}
\end{figure}

\subsection{Intra-Block Data Aggregation}

\label{section:3.2}

To improve data locality and cache efficiency, CB-SpMV employs an intra-block data aggregating strategy, as shown in Fig.\ref{fig:6} to pack and compress data within sub-blocks tightly. This approach leverages the independence of non-zero element coordinates within each sub-block after matrix partitioning, treating all sub-block data as a single unit for fast and unified data packaging and transfer.

First, \textbf{coordinate compression} is utilized to reduce memory consumption by encoding the row and column indices (\texttt{row\_idx} and \texttt{col\_idx}) of 16×16 sub-blocks into a compact format. Each index requires only 4 bits, and the two indices are combined into a single \texttt{uint\_8} type using bitwise operations, as depicted in Fig.\ref{fig:6}. This approach is extended to other sparse formats, except for dense formats where index storage is unnecessary. Second, a \textbf{virtual pointer} (VP) mechanism is introduced to enable efficient data transfer of mixed sub-block data. The VP points to a contiguous memory region on the GPUs, allocated as \texttt{char} data. Then, sub-block data of different types and precisions is aggregated into a single uniform-format sequence on the CPU, which is then transferred to the GPU in a single operation. Access on the GPU is performed via pointer offsets based on the VP and the number of non-zero elements in each sub-block, ensuring both efficiency and simplicity. Finally, since misalignment may lead to incorrect computations when the GPU accesses multiple bytes simultaneously. As shown in Fig.\ref{fig:6}(b), a \textbf{padding strategy} is applied to address potential memory misalignment caused by GPU's data alignment constraints. \enlargethispage{-15pt}

\subsection{Computational Adaptation}

Although CB-SpMV can have higher data locality after intra-block data aggregation, we also face the under-utilize challenge to implement SpMV: high sparsity sub-blocks cannot fully occupy the threads within a warp, while dense sub-blocks stored in the COO format incur memory overhead and require atomic operations, reducing parallel efficiency. To address these challenges, CB-SpMV introduces a computational adaptation strategy tailored to sub-block sparsity.\enlargethispage{-15pt}

\subsubsection{Sparse Sub-blocks}

For sparse sub-blocks, a block-aware \textbf{column aggregation} strategy is proposed. As shown in Fig.\ref{fig:5}(b), columns with all-zero elements in a sub-block are removed, and the remaining columns are shifted forward. Two arrays, \texttt{restore\_cols} and \texttt{cols\_offset}, map aggregated columns back to their original indices and record the number of non-zero columns, respectively. This ensures that each non-zero sub-block contains at least 16 non-zero elements, mitigating sparsity. 
However, block-aware column aggregation introduces stridden memory access to the $x$ vector, as aggregated column indices may no longer be contiguous. This prevents pre-loading $x$ into shared memory, slightly increasing global memory access. To balance trade-offs, we set a threshold, $th_0$, to determine if column aggregation is applied based on the sparsity of sub-blocks. As we mentioned, over 90\% of sub-blocks in most matrices have lower than 32 non-zero values, which we call these sub-blocks are super-sparse sub-blocks. Considering the rare cases where the proportion of super-sparse sub-blocks to the total number of blocks is low after matrix partitioning, storing the $x$ data of these sub-blocks in shared memory mitigates the performance gap. Consequently, column aggregation is not applied to these matrices. Thus, in our work, $th_0$ is set to a relatively small value of $0.15$.

\subsubsection{Dense Sub-blocks}

For excessively dense sub-blocks, CB-SpMV employs a \textbf{format selection} strategy, storing sub-blocks in one of three formats: COO for low-density blocks(the number of non-zero elements in a block is less than $th_1$), Dense for high-density blocks(the number of non-zero elements in a block is more than $th_2$), and CSR for intermediate sparsity. Refer to previous work TileSpMV\cite{niu2021tilespmv}, $th_1$ and $th_2$ are set to 32 and 128, respectively. This approach minimizes branching overhead while optimizing intra-block computation using the \texttt{shfl} function to enhance warp-level performance, as Alg.\ref{alg:3} and Alg.\ref{alg:4} illustrate. 

\subsection{Inter-thread-block Load Balance}

\begin{figure}[t!]
    \centering
    \includegraphics[width=1\linewidth]{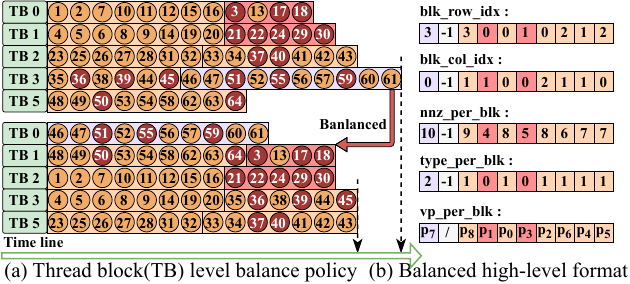}
    \caption{\small{\textbf{Block level load balancing policy and high-level structure after balancing}}}
    \label{fig:8}
\end{figure}

To address load imbalance among thread blocks, we proposed an inter-block exchange strategy to adjust the sub-block allocation for each thread block while maintaining the total number of sub-blocks of each thread block to be approximately equal. 
This involves reorganizing the metadata arrays \texttt{blk\_row\_idx}, \texttt{blk\_col\_idx}, \texttt{nnz\_per\_blk}, \texttt{type\_per\_blk},\enlargethispage{12pt}
 and \texttt{vp\_per\_blk} by priority queue(pq), as shown in Fig.\ref{fig:8}. The figure shows our load balancing example for the matrix in Fig.\ref{fig:5}, with two warps per thread block(TB), and the simplified pseudocode is shown in Alg.\ref{alg:2}. 

\begin{algorithm}[h]
\caption{A pseudocode of Thread Block Load Balance.}
\small
\label{alg:2}
\begin{algorithmic}[1]
        \State \textcolor{commentgreen}{//The blk\_idx\_array contains 3 items: ori,end,nnz.}
        \State \textbf{parallel sort}(blk\_idx\_array, \texttt{cmp\_nnz});
        \State \textcolor{commentgreen}{//The pq contains 3 items: loads(min-heap), tb\_id, warps.}
	\For{$i=0$ to $blk\_num-1$ \textbf{in parallel}}
            \State $pqtop \gets pq.top(), \text{then } pq.pop()$
            \State \textcolor{commentgreen}{// Mapping sub-block to target thread block}
	    \State $blk\_idx\_array[i].end \gets pqtop.tb\_id\times8+pqtop.warps$
            \State $pqtop.loads \gets pqtop.loads + blk\_index\_array[i].nnz$
            \State $pqtop.warps \gets pqtop.warps + 1$
            \If{$pqtop.warps < 8$}
                \State $pq.push(pqtop)$
            \EndIf
        \EndFor
        \State \textbf{parallel sort}(blk\_idx\_array, \texttt{cmp\_end});
        \For{$i=0$ to $blk\_num-1$ \textbf{in parallel}}
            \State $vp\_per\_blk[i] \gets vp\_per\_blk\_old[blk\_idx\_array[i].ori]$
            \State \textcolor{commentgreen}{//The high-level array (eg.blk\_row\_idx,...) op similarly}
        \EndFor
\end{algorithmic}
\end{algorithm}

\vspace{40pt}

This approach leverages two key innovations:
\begin{enumerate}[left=0pt]
    \item The high-level independent COO structure enables flexible rearrangement of sub-block indices while ensuring efficient localization within the global matrix.
    \item All data in each sub-block is stored contiguously, allowing the virtual pointer (VP) to be directly retrieved for efficient computation access.
\end{enumerate}

Compared to previous methods that use CSR as the high-level format and store sub-block data non-contiguously, this strategy achieves superior load balancing by enabling dynamic and efficient sub-block remapping without altering the total number processed per thread block.

\subsection{Kernel Implementation}

To accommodate the proposed optimization measures, the SpMV kernel's computation logic and algorithm have been redesigned to fully exploit the CB-SpMV format's advantages.

\begin{algorithm}[h]
\caption{A pseudocode of CB-SpMV with COO.}
\label{alg:3}
\small
\begin{algorithmic}[1]
	\For{\text{each block matrix} $A_k$ \text{in the matrix} $A$ \textbf{in parallel}}
	    \State $blk\_nnz$ $\gets$ the nnz of $A_k$
	    \State $VP$ $\gets$ the virtual pointer of $A_k$
            \State \textcolor{commentgreen}{// Unpacking data}
            \State $blk\_row\_idx \gets \text{the row block index of } A_k$
            \State $padding \gets (blk\_nnz \times \text{size}(Idx)) \bmod \text{size}(Val)$
            \State $padding \gets padding \,\textbf{?}\, \text{size}(Val)-padding \,\textbf{:}\, 0$
            \State $coo\_idx \gets VP$
            \State $coo\_val \gets VP + blk\_nnz \times \text{size}(Idx) + padding$
            \For{$i=0$ \textbf{to} $blk\_nnz$ \textbf{in parallel} }
                    \State $row\_idx \gets blk\_coo\_idx[i] \,\&\, 15$
                    \State $col\_idx \gets blk\_coo\_idx[i] \gg 4$
                    \State $y\_idx \gets blk\_row\_idx \times \text{BLK\_SIZE} + row\_idx$
                    \If {column aggregation is not used}
                        \State \textcolor{commentgreen}{// Corresponding x is preloaded into sm.}
                        \State $\text{atomicADD}(y_{y\_idx},coo\_val[i] \times s\_x_{col\_idx})$ 
                    \Else
                        \State \textcolor{commentgreen}{//Obtain the global pointer of x.}
                        \State $offset \gets cols\_offset[blk\_row\_idx]+col\_idx $
                        \State $col\_idx\_ori \gets restore\_cols[offset]$
                        \State $\text{atomicADD}(y_{y\_idx},coo\_val[i] \times d\_x_{col\_idx\_ori})$
                    \EndIf
            \EndFor
        \EndFor
\end{algorithmic}
\end{algorithm}

For sub-blocks stored in the COO format, a warp of 32 threads is allocated to process all non-zero elements in one sub-block. Even if the number of non-zero elements is less than 32, the column aggregation strategy ensures that each sub-block contains at least 16 elements, improving hardware utilization to at least 50\%. This marks a significant improvement over traditional approaches, which often leave most threads idle in sparse sub-blocks. The packed data for each sub-block is parsed using the metadata arrays \texttt{nnz} and VP to extract \texttt{row\_idx}, \texttt{col\_idx}, and \texttt{val}. The computation results are then partially accumulated into the corresponding \texttt{y} vector using the \texttt{atomicAdd} operation, ensuring correctness across concurrent threads. 

\begin{algorithm}[h]
\caption{A pseudocode of CB-SpMV with Dense.}
\label{alg:4}
\small
\begin{algorithmic}[1]
	\For{\text{each block matrix} $A_k$ \text{in the matrix} $A$ \textbf{in parallel}}
	    \State $blk\_nnz \gets \text{nnz of } A_k$
	    \State $VP \gets \text{virtual pointer of } A_k$
        \State $blk\_row\_idx \gets \text{row block index of } A_k$
        \State $dense\_val \gets VP$
        \For{$tid=0$ \textbf{to} $warpsize-1$ \textbf{in parallel}}
            \State $sum \gets 0$
            \State $row\_ft \gets (tid < \text{BLK\_SIZE}) ? tid : tid - \text{BLK\_SIZE}$
            \State $start \gets (tid < \text{BLK\_SIZE}) ? 0 : \text{BLK\_SIZE}/2$
            \State $end \gets (tid < \text{BLK\_SIZE}) ? \text{BLK\_SIZE}/2 : \text{BLK\_SIZE}$
            \For{$col=start$ \textbf{to} $end$}
                \State $val \gets dense\_val[row\_ft \times \text{BLK\_SIZE} + col]$
                \If{column aggregation is not used}
                    \State $sum += val \times s\_x_{col}$
                \Else
                    \State $offset \gets cols\_offset[blk\_row\_idx] + col$
                    \State $col\_idx\_ori \gets restore\_cols[offset]$
                    \State $sum += val \times d\_x_{col\_idx\_ori}$
                \EndIf
            \EndFor
            \If{$tid < \text{BLK\_SIZE}$}
                \State $sum \gets \text{\_\_shfl\_xor\_sync}(\text{0xffffffff}, sum, \text{BLK\_SIZE}/2)$
                \State \textcolor{commentgreen}{\Comment{Aggregate partial results within warp}}
            \EndIf
        \EndFor
        \For{$tid=0$ \textbf{to} $\text{BLK\_SIZE}$ \textbf{in parallel}}
            \State $y\_idx \gets blk\_row\_idx \times \text{BLK\_SIZE} + tid$
            \State $\text{atomicAdd}(y_{y\_idx}, sum)$
        \EndFor
	\EndFor
\end{algorithmic}
\end{algorithm}

For sub-blocks stored in CSR or Dense formats, the kernel logic is optimized to leverage GPU-specific instructions. Since 32 threads collaboratively compute 16 \texttt{y} vector elements, the unpacking process is followed by a warp-level reduction using the \texttt{shfl} instruction. This operation minimizes access to shared or global memory, enhancing computational efficiency. The dense format further benefits from optimized memory coalescing due to its regular structure, reducing cache misses.
Alg.\ref{alg:3} and Alg.\ref{alg:4} provide pseudocode for the redesigned COO and Dense kernels. These kernels are categorized based on whether column aggregation is applied. For sub-blocks with column aggregation, the \texttt{x} vector is directly loaded from global memory into registers, requiring a remapping from aggregated to original column indices. Conversely, for sub-blocks without column aggregation, the corresponding portion of the \texttt{x} vector is pre-loaded into shared memory to minimize memory latency during computation. This distinction ensures that the kernel adapts dynamically to different sub-block sparsity levels, maximizing performance while maintaining simplicity. 

%% file: sections/experiment.tex
\begin{figure*}[t!]
    \centering
    \includegraphics[width=1\linewidth]{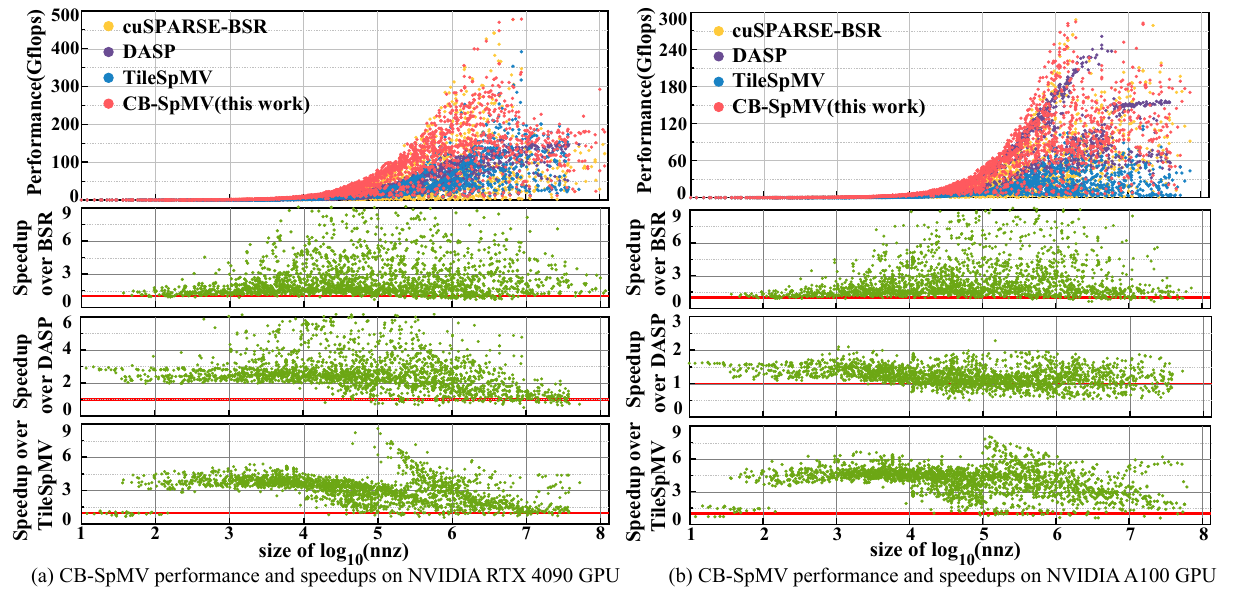}
    \caption{\small{\textbf{Performance comparison between CB-SpMV and the SOTA SpMV algorithm on RTX4090 and A100 GPUs}}}
    \label{fig:9}
\end{figure*}

\section{Evaluation}

\label{section:4}

\subsection{Experimental Setup}

The experimental platform consists of two types of NVIDIA GPUs: the NVIDIA A100 (Ampere architecture) and RTX 4090 (Ada Lovelace architecture). Both GPUs are configured with driver version 550.135 and CUDA version 12.4. On this platform, the proposed method is comprehensively compared with the following state-of-the-art approaches: The \texttt{cusparse\_bsrmv()}\cite{cusparse} kernels from cuSPARSE v12.4 for SpMV using BSR formats;  the state-of-the-art block-based SpMV method TileSpMV\cite{niu2021tilespmv}; and the latest based on CSR format and Tensor Core-accelerated SpMV method DASP\cite{lu2023dasp}. The table below lists the specifications of the tested GPUs and the algorithms. The evaluation dataset comprises 2843 matrices from the SuiteSparse Matrix Collection\cite{davis2011university}.

\begin{table}[t]
  \centering
  \caption{The two GPUs and four algorithms evaluated.}
  \begin{tabular}{|p{0.55\linewidth}|p{0.35\linewidth}|} 
    \hline
    \textbf{Two NVIDIA GPUs} & \textbf{Four Algorithms} \\ \hline
    (1) NVIDIA A100 (Ampere), 6912 CUDA cores @ 1410 MHz, 40 GB, B/W 1.56 TB/s 
    \newline 
    (2) NVIDIA RTX4090 (Ada Lovelace), 16384 CUDA cores @ 2520 MHz, 24GB, B/W 1.01 TB/s
    & 
    (1) cuSPARSE \cite{cusparse}
    \newline
    (2) TileSpMV  \cite{niu2021tilespmv}
    \newline 
    (3) DASP \cite{lu2023dasp}
    \newline 
    \textbf{(4) CB-SpMV  \newline (this work)} \\ \hline
  \end{tabular}
\end{table}

\subsection{SOTA Technology Comparison}

In this subsection, we evaluate the performance of CB-SpMV (with \texttt{val} in FP64 format) against the SOTA block-based SpMV algorithms (e.g., cuSPARSE-BSR\cite{cusparse}, TileSpMV\cite{niu2021tilespmv}) and other recent SpMV methods (DASP\cite{lu2023dasp}), focusing on Gflop/s (Gflops) as the primary performance metric. 
We run each kernel 1000 times to obtain average Gflops values. For cuSPARSE-BSR, we use the best-performing block size from 2×2, 4×4, 8×8, and 16×16 while for TileSpMV and our work, we set the block size as 16x16.
Additionally, we analyze the impact of the intra-subblock data aggregation strategy on L1 and L2 cache hit rates.\vspace{-5pt}

\subsubsection{Performance Comparison}

As shown in Fig.\ref{fig:9}, CB-SpMV significantly outperforms BSR on both GPUs. On the RTX 4090, CB-SpMV achieves an average speedup of 2.95× and a maximum speedup of 37.54× (on \texttt{TSOPF\_FS\_b39\_c30}). On the A100, the average and maximum speedups are 2.99× and 54.27× (on \texttt{boyd1}). These improvements stem from CB-SpMV’s use of sparse formats for low-level sub-blocks, avoiding the dense storage overhead of BSR and reducing underutilization in sparse sub-blocks. CB-SpMV also consistently outperforms TileSpMV. For smaller matrices ($nnz$ less than $10^5$), the primary gains arise from enhanced data locality via intra-subblock data aggregation. For larger matrices, strategies like column aggregation, format selection, and load balancing further boost performance. On the RTX 4090 and A100, CB-SpMV achieves average speedups of 3.06× and 3.95×, with maximum speedups of 8.56× (on \texttt{piston}) and 10.34× (on \texttt{rgg\_n\_2\_21\_s0}). 

\begin{figure*}[t]
    \centering
    \includegraphics[width=1\linewidth]{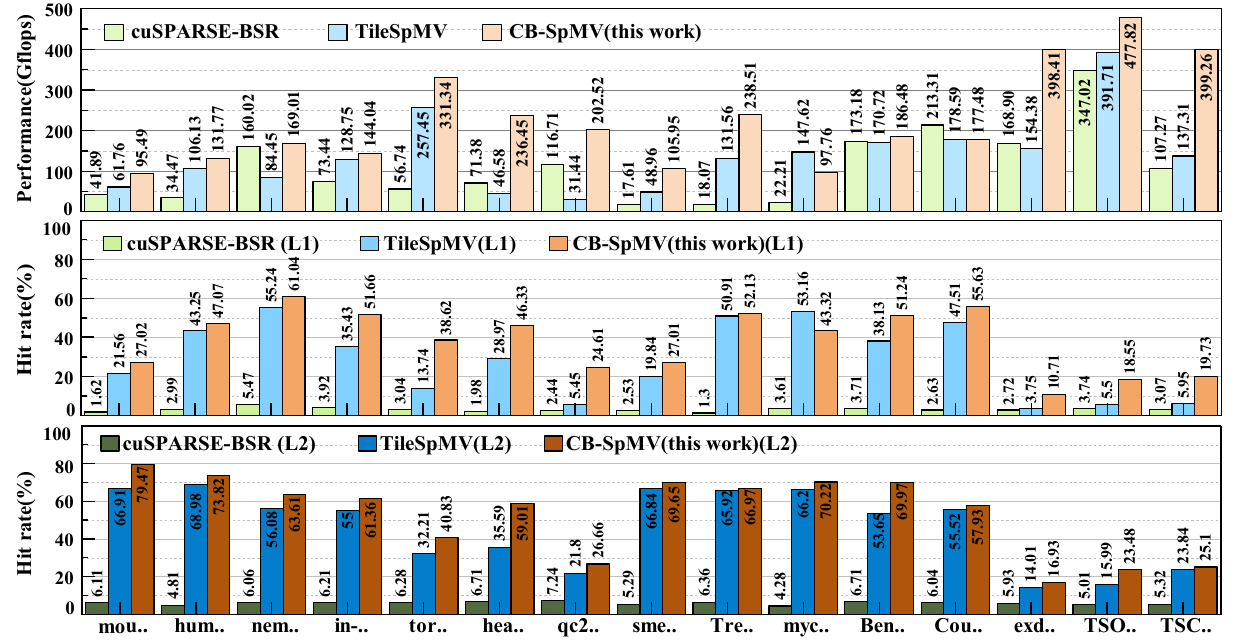}
    \caption{\small{\textbf{Performance comparison of 15 typical sparse matrices on RTX 4090 GPU and L1 \& L2 Cache hit ratio}}}
    \label{fig:10}
\end{figure*}

The comparison with DASP, which leverages Tensor Core designs, is particularly noteworthy. On the RTX 4090 and A100 GPUs, CB-SpMV achieves average speedup factors of 2.76× and 1.21×, respectively. The reduced performance advantage of the A100 is primarily due to its design as a high-performance computing GPU, offering significantly higher FP64 Tensor Core computational capabilities compared to the RTX 4090, which is optimized for graphics processing. Nevertheless, despite DASP’s reliance on the powerful Tensor Core hardware, CB-SpMV still achieves approximately 20\% speedup on the A100. This is likely because DASP’s data layout strategy fails to fully exploit the potential of Tensor Cores, as discussed in Section \ref{section:1}. Addressing this issue will be a key focus of our future work.

\subsubsection{Cache Hit Rate Comparison}

\vspace{-5pt}
To vividly illustrate the improvements in data locality achieved by sub-block data aggregation and the associated strategies, we selected 15 representative matrices, as shown in the table \ref{tab:2}. Using the Nsight Compute tool on the RTX 4090, we analyzed parameters such as \texttt{l1tex\_\_t\_sector\_hit\_rate} for L1 Cache hit rate and \texttt{lts\_\_t\_request\_hit\_rate} for L2 Cache hit rate. These metrics were used to evaluate the Gflops and average L1 and L2 cache hit rates for the three methods mentioned earlier when applied to these 15 matrices. 

\renewcommand{\arraystretch}{1.5} 
\begin{table}[t!]
\centering
\caption{Information of the 15 Representative Matrices.}
\label{tab:2}
\vspace{-5pt}
\begin{tabular}{|>{\centering\arraybackslash}m{2cm}|>{\centering\arraybackslash}m{1.7cm}|>{\centering\arraybackslash}m{2cm}|>{\centering\arraybackslash}m{1.3cm}|}
\hline
\textbf{Name}   & \textbf{Plot}  & \textbf{Size} ($m \times n$)        & $nnz$     \\ \hline
mouse\_gene   & \raisebox{-0.3\height}{\includegraphics[height=20pt]{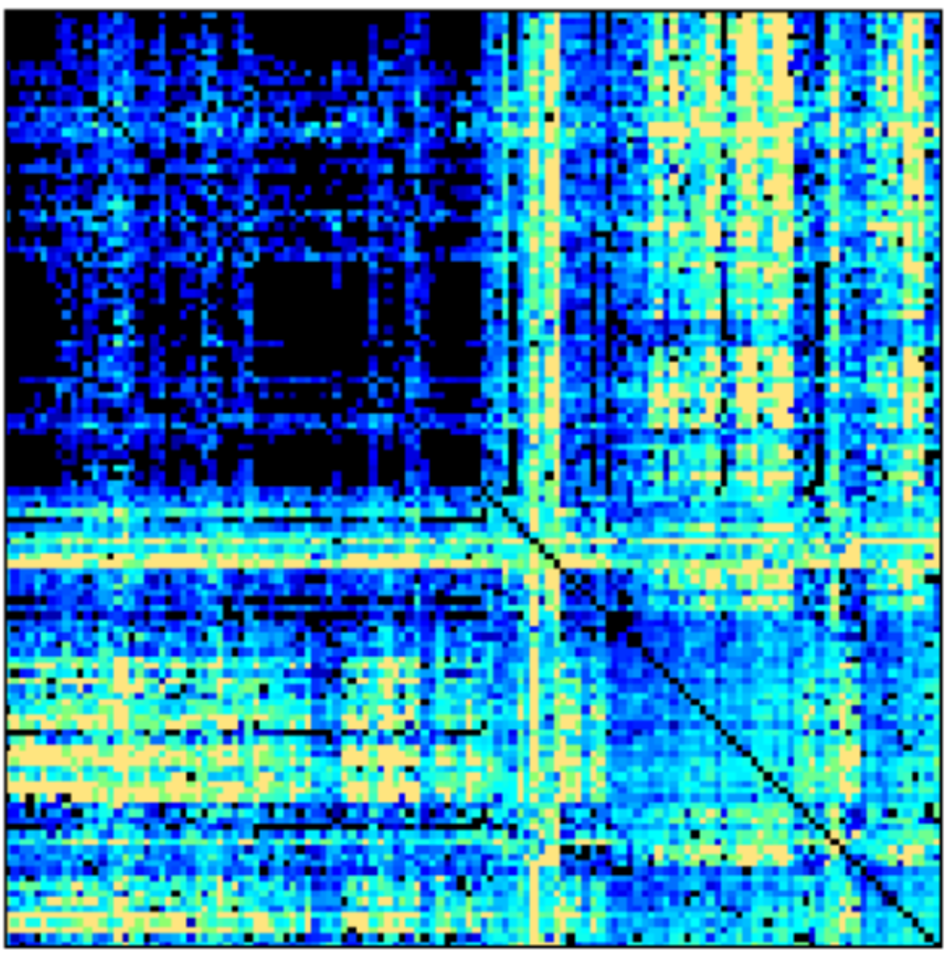}}         & 45101×45101     & 28967291   \\ \hline
human\_gene1  & \raisebox{-0.3\height}{\includegraphics[height=20pt]{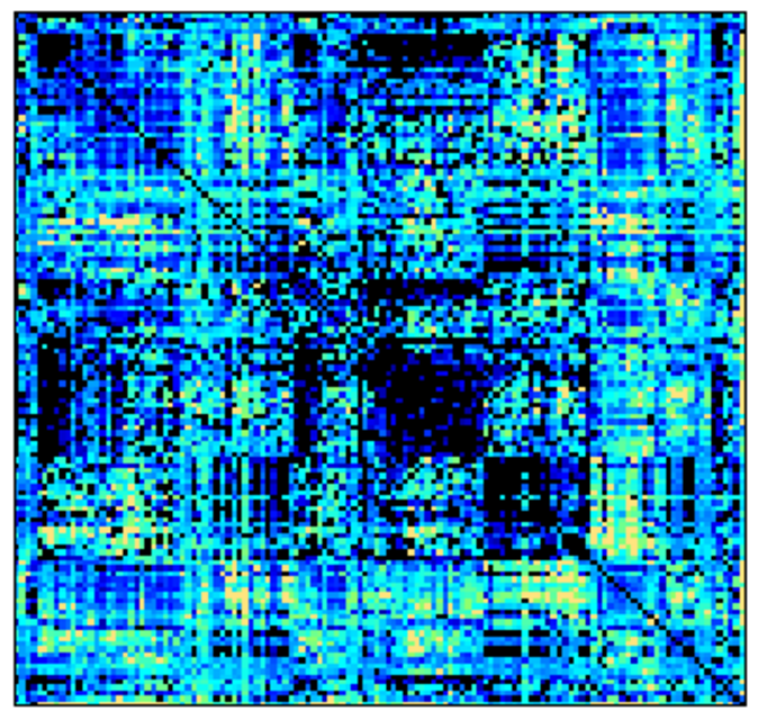}}         & 22283×22283     & 24669643   \\ \hline
nemeth07      & \raisebox{-0.3\height}{\includegraphics[height=20pt]{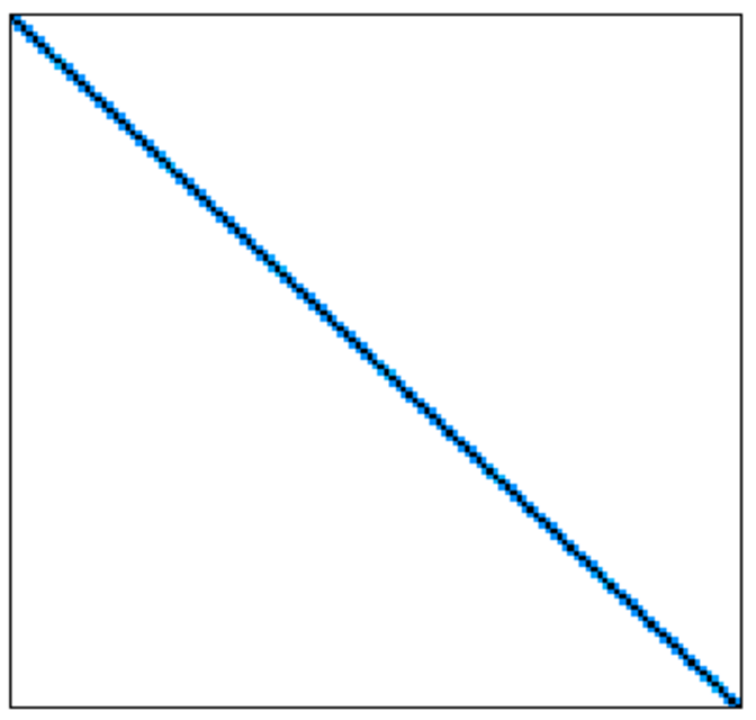}}         & 9506×9506       & 394812   \\ \hline
in-2004       & \raisebox{-0.3\height}{\includegraphics[height=20pt]{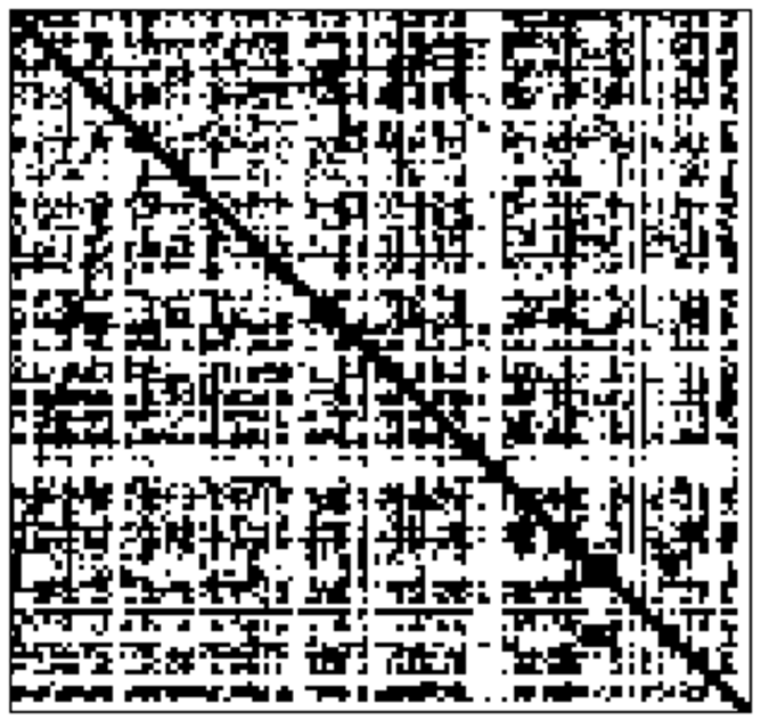}}         & 1382908(m=n)    & 16917053   \\ \hline
torso1        & \raisebox{-0.3\height}{\includegraphics[height=20pt]{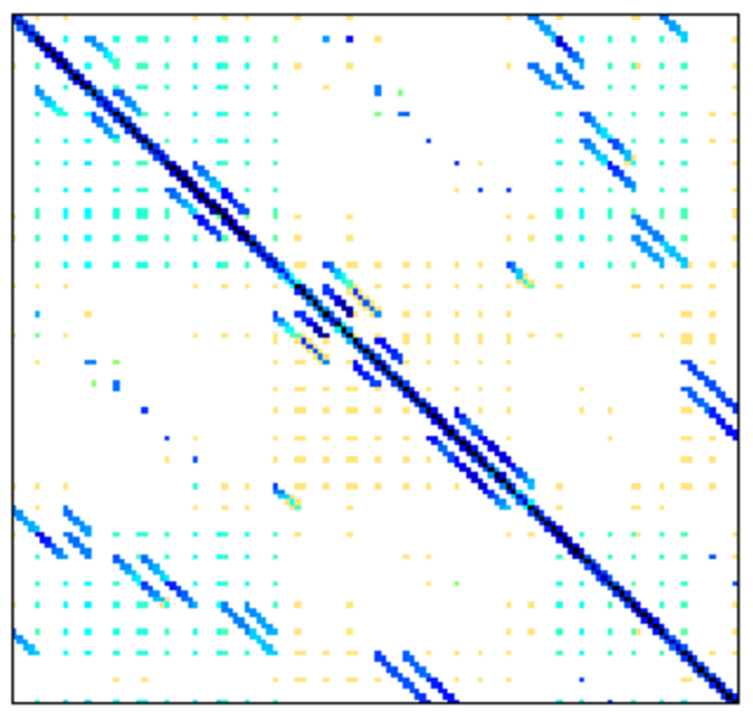}}         & 116158×116158   & 8516500   \\ \hline
heart2        & \raisebox{-0.3\height}{\includegraphics[height=20pt]{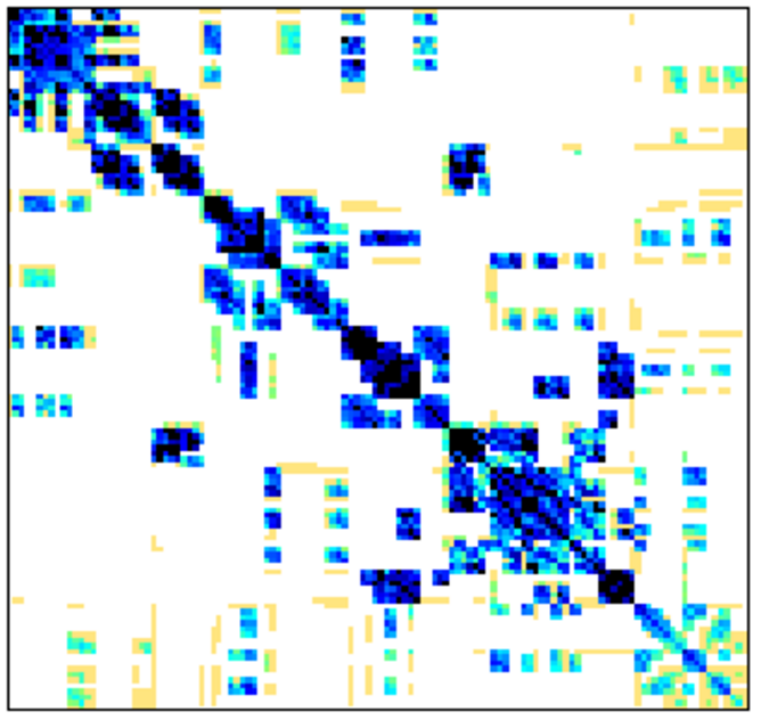}}         & 2339×2339       & 682797   \\ \hline
qc2534        & \raisebox{-0.3\height}{\includegraphics[height=20pt]{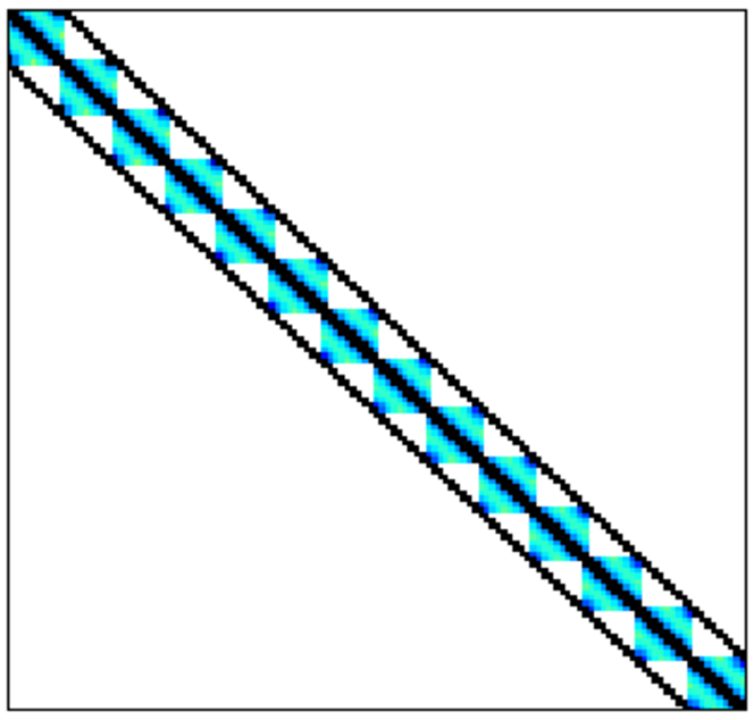}}         & 2534×2534       & 463360  \\ \hline
sme3Da        & \raisebox{-0.3\height}{\includegraphics[height=20pt]{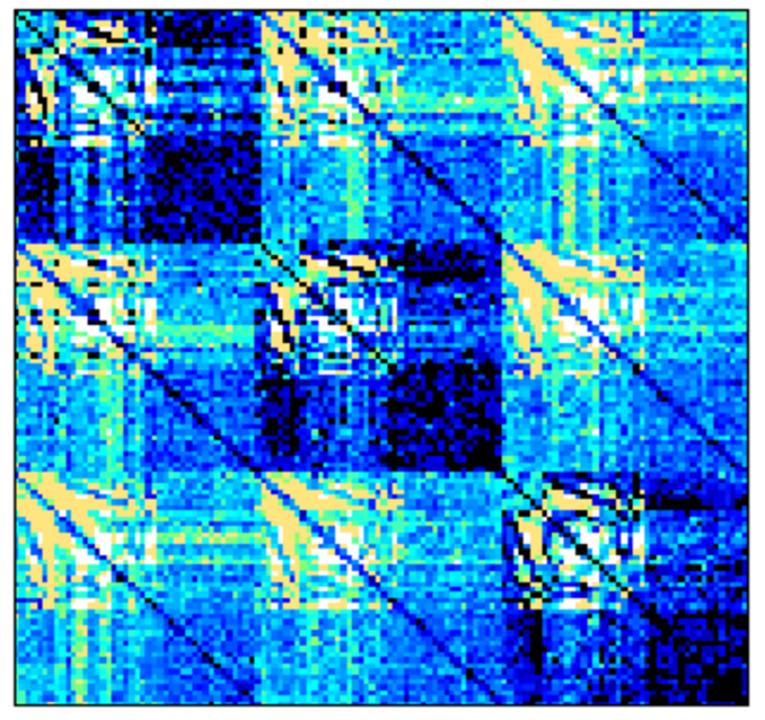}}         & 12504×12504     & 874887  \\ \hline
Trec14        & \raisebox{-0.3\height}{\includegraphics[height=10pt]{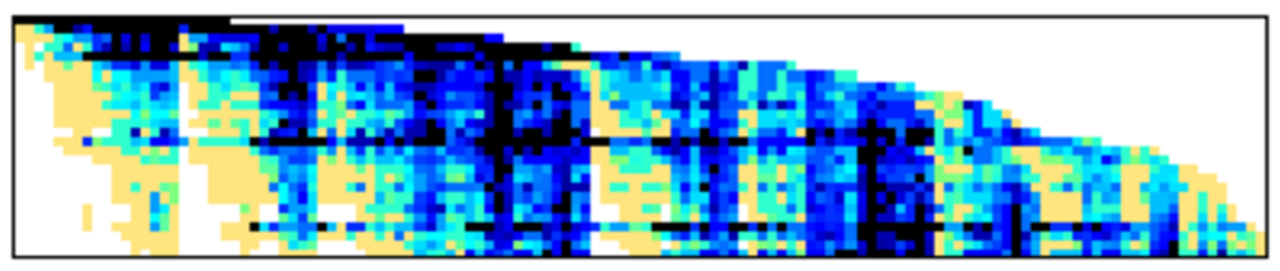}}         & 3159×15905      & 2872265  \\ \hline
mycielskian15 & \raisebox{-0.3\height}{\includegraphics[height=20pt]{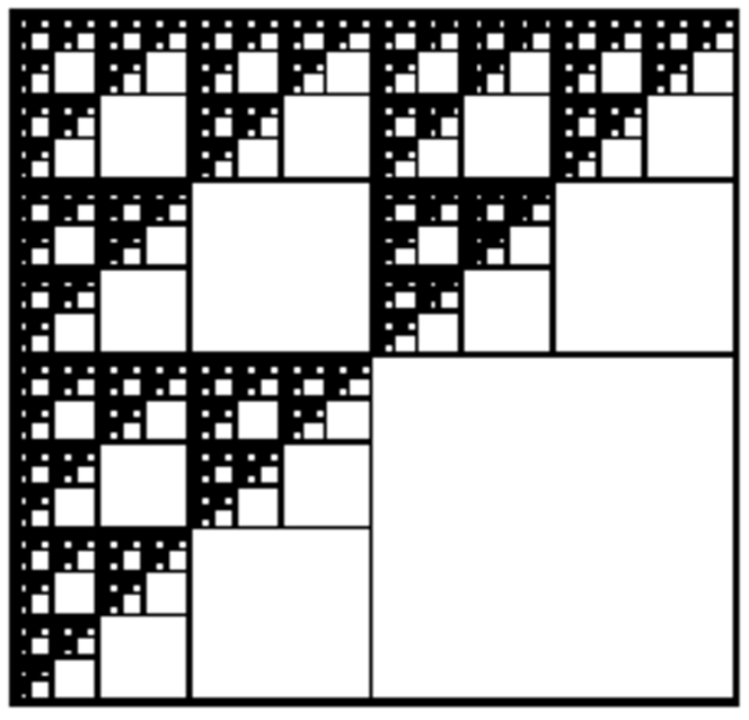}}        & 24575×24575     & 11111110  \\ \hline
BenElechi1    & \raisebox{-0.3\height}{\includegraphics[height=20pt]{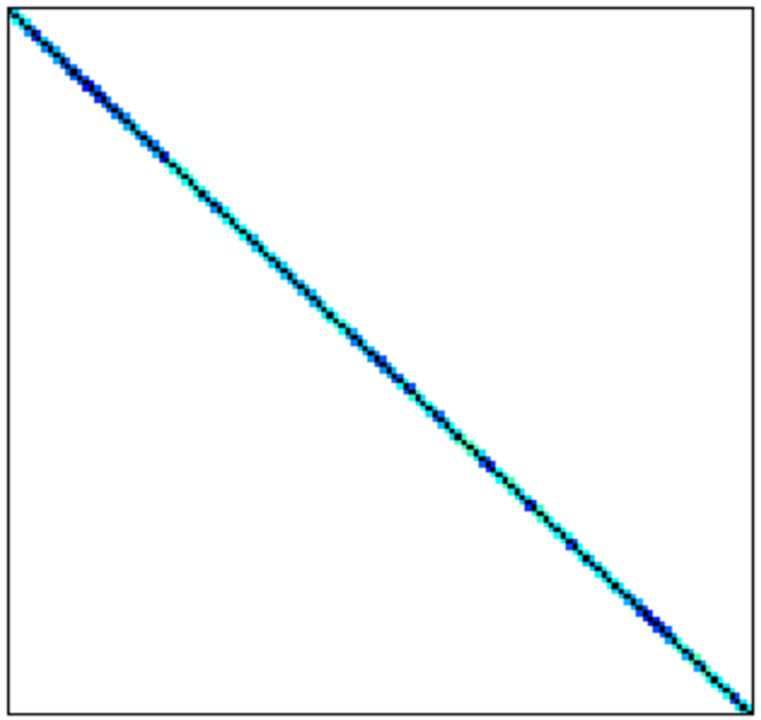}}        & 245874×245874   & 13150496  \\ \hline
CoupCons3D    & \raisebox{-0.3\height}{\includegraphics[height=20pt]{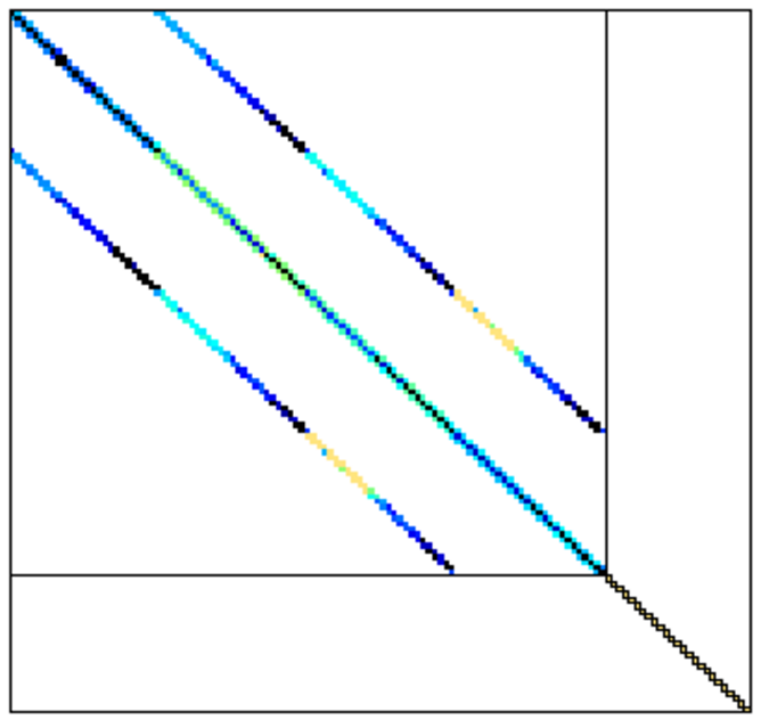}}        & 416800×416800   & 22322336  \\ \hline
exdata\_1      & \raisebox{-0.3\height}{\includegraphics[height=20pt]{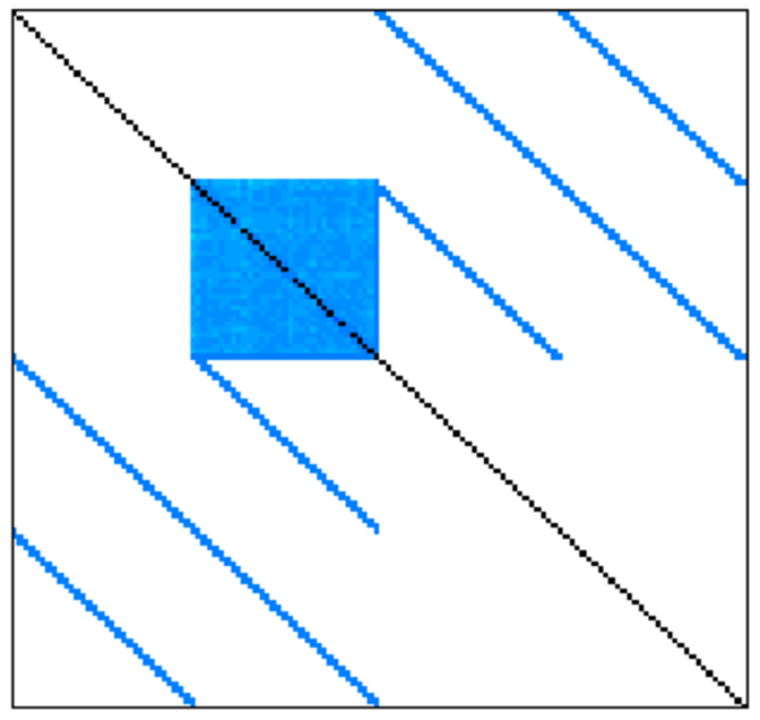}}        & 6001×6001       & 2269501  \\ \hline
TSOPF\_RS\_b6.. & \raisebox{-0.3\height}{\includegraphics[height=20pt]{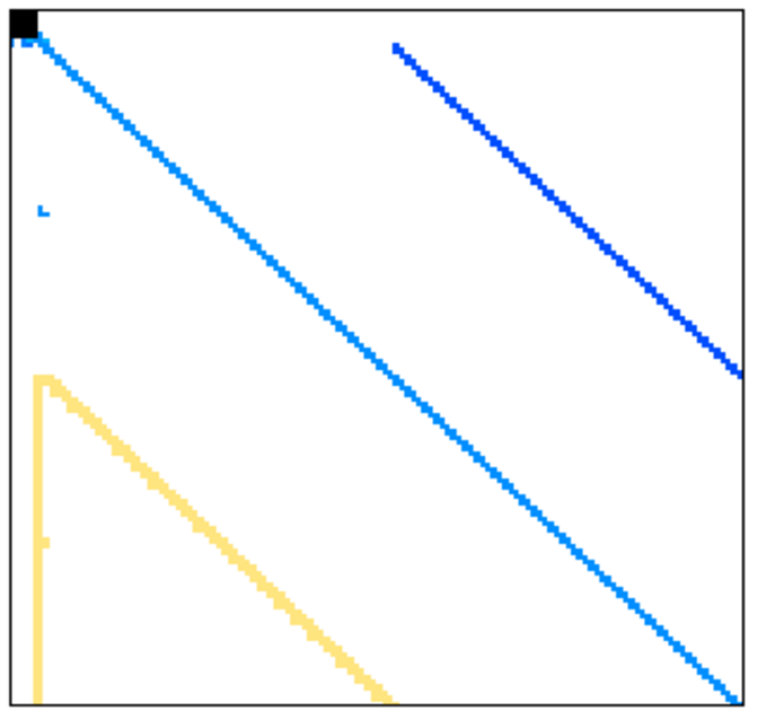}}  & 35696×35696     & 8781949  \\ \hline
TSC\_OPF\_1047 & \raisebox{-0.3\height}{\includegraphics[height=20pt]{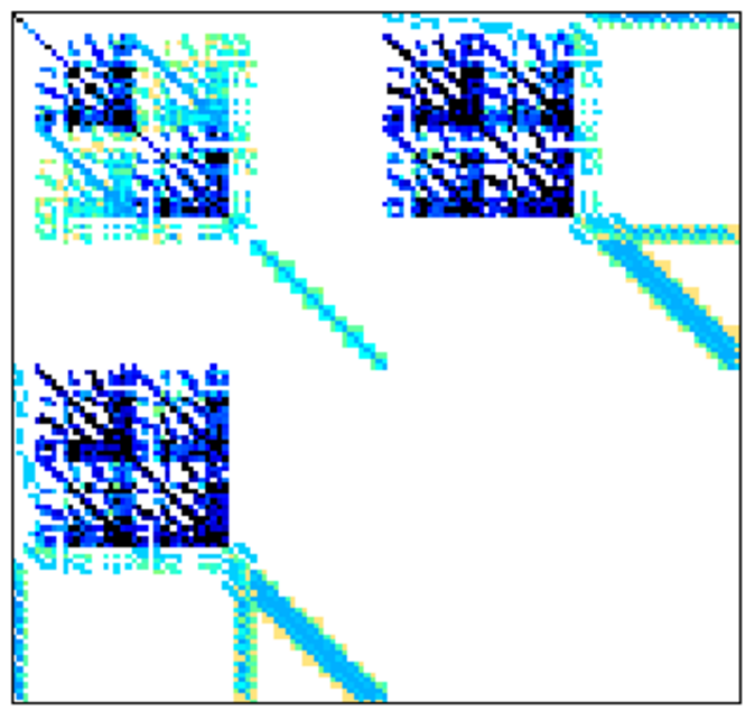}}        & 8140×8140     & 2016902  \\ \hline
\end{tabular}
\end{table}

\begin{figure*}[t!]
    \centering
    \includegraphics[width=1\linewidth]{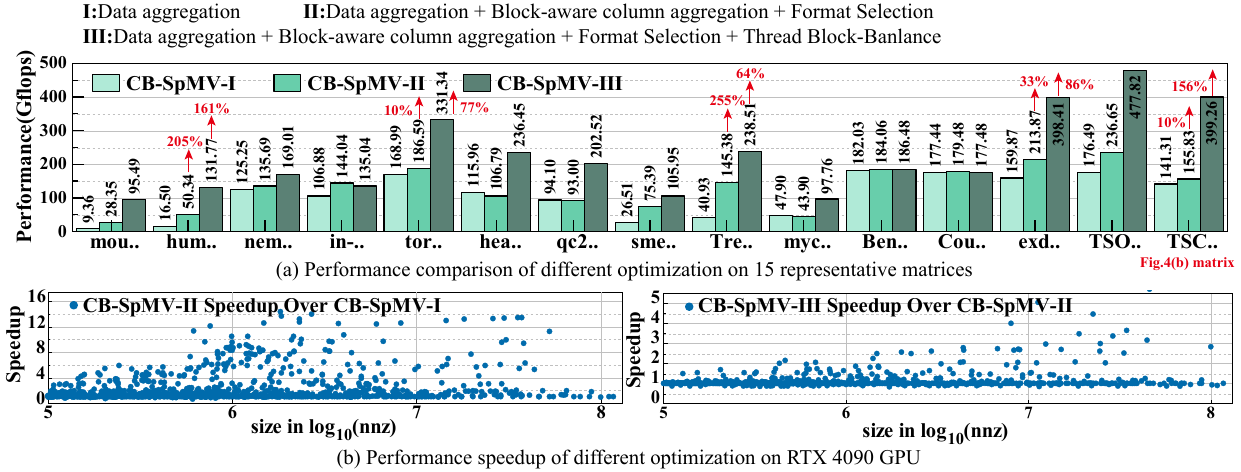}
    \caption{\small{\textbf{Performance comparison of 15 typical sparse matrices using different optimization measures using CB-SpMV on RTX 4090 GPUs.}}}
    \label{fig:12}
\end{figure*}

As shown in Fig.\ref{fig:10}, CB-SpMV achieves L1 cache hit rates 14.4× and 1.93× higher than BSR and TileSpMV, respectively, and L2 hit rates 5.89× and 1.19× higher. Additionally, matrices with more dense sub-blocks, such as \texttt{mouse\_gene} and \texttt{sme3Da}, exhibit higher L2 cache hit rates. This is because dense sub-blocks, after aggregation, often require larger storage spaces, leading to more frequent evictions and reloads in the L1 cache. In contrast, matrices with more sparse sub-blocks, such as \texttt{nemeth07} and \texttt{CoupCons3D}, tend to have higher L1 cache hit rates. This is because sparse sub-blocks contain fewer data elements, allowing the L1 cache to accommodate more of their content. These patterns underscore the adaptability of CB-SpMV in improving data locality across matrices with varying structural characteristics.


\subsection{Ablation Performance Analysis}

To address the challenges outlined in Section \ref{section:2}, we proposed targeted optimization strategies in Section \ref{section:3}. This section evaluates their individual and combined impacts on CB-SpMV’s performance, focusing on 15 representative matrices with $nnz > 100,000$, as preprocessing selectively applies these optimizations based on matrix characteristics.

Fig.\ref{fig:12} highlights the contributions of each optimization strategy. Compared to using only intra-block data aggregation (CB-SpMV-I), the column aggregation and format selection strategy achieves an average speedup of 2.22×, and the thread block-level load balancing strategy provides an additional 1.09× improvement compared to using sub-block data aggregation, column aggregation and format selection (CB-SpMV-II). When combined, these strategies deliver an average performance gain of 2.37×. The column aggregation and format selection strategy addresses issues in matrices with extreme sub-block sparsity or density. For example, in \texttt{human\_gene1}, this optimization mitigates reduced parallelism caused by dense sub-blocks, achieving a 205\% improvement. Conversely, for \texttt{exdata\_1}, it enhances warp utilization in sparse sub-blocks, delivering a 33\% gain. The thread block-level load balancing strategy effectively distributes the workload among thread blocks, resolving imbalances in matrices with mixed dense and sparse regions, such as \texttt{torso1} and \texttt{exdata\_1}, achieving 77\% and 86\% improvements, respectively.


\subsection{Overhead Analysis}

In this section, we evaluate the advantages and disadvantages of the proposed method in terms of storage overhead and preprocessing time and compare it with other sparse matrix storage formats.

\subsubsection{storage overhead}

\begin{figure}[t!]
    \centering
    \includegraphics[width=1\linewidth]{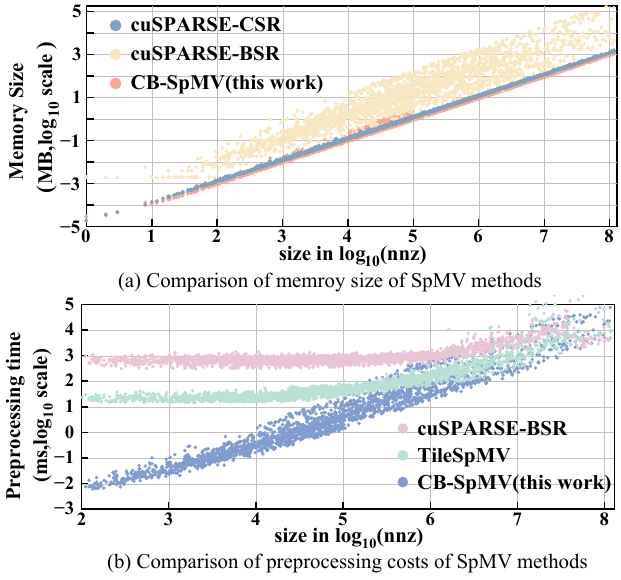}
    \caption{\small{\textbf{Comparison of memory usage and preprocessing time across different SpMV storage formats.}}}
    \label{fig:13}
\end{figure}

To evaluate storage overhead, we modeled the space usage of CSR, BSR, and CB-SpMV under simplified assumptions: position information is stored as \texttt{int32}, and numerical values are stored as \texttt{FP64}. For a matrix of size $m \times n$ with $nnz$ non-zero elements, the CSR format requires $(m+1)\times 4 + nnz\times 4 + nnz\times 8 \text{B}$. For the BSR format, assuming a block size of $16 \times 16$, with $nnzb$ non-zero sub-blocks and $blk\_m$ row blocks, the required storage is $256 \times 8 \times nnzb + (blk\_m+1) \times 4 + nnzb \times 4 \text{B}$. For CB-SpMV, where each sub-block uses COO storage, the storage overhead is $nnzb \times (4+4+4+1+8) + nnz \times (1+8) \text{B}$. Column aggregation costs are excluded, as the reduction in block count offsets high-level storage costs. As shown in Fig.\ref{fig:13}(a), CB-SpMV achieves storage efficiency comparable to CSR, thanks to its compressed sub-block storage and smaller indices. In contrast, BSR incurs significantly higher overhead due to storing numerous \texttt{FP64} zero elements within sub-blocks. This demonstrates that CB-SpMV effectively balances storage efficiency and computational adaptability.

\subsubsection{preprocessing time overhead}

The preprocessing time required to convert COO matrices into the formats used by various algorithms is shown in Fig.\ref{fig:13}(b). CB-SpMV consistently outperforms TileSpMV and cuSPARSE-BSR for matrices with $nnz$ less than $10^6$. For matrices with $nnz$ greater than one million, CB-SpMV remains comparable to those of TileSpMV and BSR. Despite the modest increase in preprocessing time for larger matrices, CB-SpMV remains competitive, particularly for iterative solvers that require repeated SpMV operations. The preprocessing overhead is a small trade-off for the significant performance improvements achieved during computation.

%% file: sections/related_work.tex
\section{Related Work}

\label{section:5}

There has been extensive prior research on SpMV; here, we focus only on reviewing works relevant to this study.

\textbf{Block-based Methods:}In recent years, with the rise of block and tensor core, block-based approaches about GPUs have garnered significant attention in these studies\cite{labini2022blocking, lu2020efficient, yang2018parallel, niu2021tilespmv, niu2022tilespgemm, eberhardt2016optimization, ashari2014efficient, bulucc2009parallel, vuduc2005fast, navarro1996block, o1990block, borvstnik2014sparse, bi2023efficiently}. Researchers like Labini and Bernaschi \cite{labini2022blocking}  leveraged reordering and blocking techniques and dense accelerators such as NVIDIA Tensor Cores to achieve efficient sparse matrix multiplications. Buluç\cite{buluc2008representation, bulucc2009parallel, bulucc2011reduced, bulucc2012parallel} et al. introduced the CSB format to exploit block layouts and cache locality, which Martone further improved using recursive methods. Yan\cite{yan2014yaspmv} et al. developed the BCCOO format to store dense 2D blocks, and Niu\cite{niu2021tilespmv} et al. extended this idea with TileSpMV by supporting seven internal sub-block structures. Compared to these studies, our approach aggregates and packs data within each sub-block, treating each as an independent unit with tightly adjacent data. This method achieves better memory locality among block-based algorithms, significantly improving the cache hit rate for blocked matrix computations. Beyond advancing the theoretical understanding of block-based sparse matrix SpMV, this approach also opens new avenues for applying blocked matrices to other computations, such as SpMM and SpGEMM.

\textbf{Formats for SpMV:} Widely adopted strategies for accelerating SpMV computations focus on designing novel storage formats and optimizing algorithms\cite{filippone2017sparse, liu2015csr5, du2022alphasparse, gao2022taichi, tareen2024hihispmv, kim2024campus, zhang2023memory, chu2023efficient}. Among them, numerous ELL and CSR-based formats have been proposed to enhance performance. Notable examples include the HYB format combining ELL and CSR/COO\cite{bell2009implementing} and the clSpMV framework integrating multiple formats. Variants of ELL\cite{kreutzer2014unified, liu2015csr5, liang2017scale, ashari2014efficient, anzt2020load, xie2021fast} and CSR\cite{kourtis2011csx, ashari2014fast, liu2015csr5, merrill2016merge, steinberger2017globally} formats have demonstrated significant advantages, especially on GPUs. Yesil et al.\cite{yesil2020speeding} further improved data locality by separating matrices into dense and sparse regions with customized representations. Despite these advances, the irregular distribution of nonzero elements in matrices limits the efficiency of single-format solutions. Our approach segments matrices into independent sub-blocks and applies computation-specific optimizations such as column aggregation and adaptive format selection, achieving enhanced flexibility and performance across diverse matrices.

\textbf{Load Balance:} Load balancing is a critical factor in improving parallel sparse matrix-vector multiplication (SpMV) performance\cite{Filippone2017, anzt2020load, bian2021albus, Greathouse2014CSR, li2023haspmv, liu2015csr5, merrill2016merge, mi2023balancing, steinberger2017globally, osama2023programming, guo2024camlb}. The uneven distribution of nonzero elements often causes workload imbalances, reducing efficiency. Various strategies have been proposed to address this issue, including the Merge-Path algorithm for fine-grained workload decomposition\cite{merrill2016merge} and methods for balancing computation and communication loads in distributed systems\cite{mi2023balancing}. Osama\cite{osama2023programming} et al introduces an abstraction model that decouples load balancing from task execution, using a hierarchical structure of work units, tiles, and sets, along with a programmable scheduling interface, to achieve GPU static and dynamic load balancing. Additionally, thread block-level strategies such as DTC-SpMM have significantly improved sparse matrix-matrix multiplication (SpMM)\cite{fan2024dtc}. Despite these advances, achieving ideal thread block-level load balancing in SpMV remains challenging. Our proposed CB-SpMV format introduces a priority queue mechanism to adjust sub-block task assignments dynamically, significantly improves workload distribution and computational efficiency between different thread blocks, and offers new insights for optimizing sparse matrix computations.\enlargethispage{-18pt}

%% file: sections/conclusion.tex
\section{Conclusion}
\label{section:6}

In this work, we propose a novel cache-friendly SpMV computation method, CB-SpMV, which is designed to optimize data locality and computational efficiency through a block-based structure. Specifically, the matrix is partitioned into independent and uniformly sized sub-blocks, where various types of data within each sub-block are aggregated using virtual pointers to improve access patterns. To handle the challenges posed by varying sparsity levels across sub-blocks, we introduce a block-wise column aggregation strategy that consolidates sparse data and a format-adaptive selection mechanism that chooses suitable storage for each sub-block. Additionally, a thread block-level load-balancing algorithm is developed to mitigate the imbalance in processing non-zero elements across thread blocks. Experimental results on two GPUs demonstrate that our CB-SpMV method achieves higher cache hit rates and superior performance compared to state-of-the-art SpMV approaches.